\documentclass{article}

\usepackage{arxiv}
\usepackage[numbers]{natbib} 
\usepackage[utf8]{inputenc} % allow utf-8 input
\usepackage[T1]{fontenc}    % use 8-bit T1 fonts
\usepackage{hyperref}       % hyperlinks
\usepackage{url}            % simple URL typesetting
\usepackage{booktabs}       % professional-quality tables
\usepackage{amsfonts}       % blackboard math symbols
\usepackage{nicefrac}       % compact symbols for 1/2, etc.
\usepackage{microtype}      % microtypography
\usepackage{lipsum}		% Can be removed after putting your text content
\usepackage{graphicx}
\usepackage{natbib}
\usepackage{doi}
\usepackage{amsmath} 
\usepackage{xcolor}
\usepackage{float}
\usepackage{epstopdf}
\usepackage{mathtools}
\newtagform{supplementary}[S.]()
\counterwithin*{equation}{section}
\usepackage{stackengine}

\title{Inverse Diffusion Approximation for extraction of scattering and absorption coefficients in \textcolor{black}{Highly Scattering} Media.}

\author{ \href{https://orcid.org/0000-0000-0000-0000}{\includegraphics[scale=0.06]{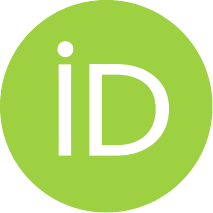}\hspace{1mm}Lingxi Li}\thanks{Use footnote for providing further
		information about author (webpage, alternative
		address)---\emph{not} for acknowledging funding agencies.} \\
	Department of Electronic and Electrical Engineering\\
	University College London\\
	London, WC1E 7JE \\
	\And
   Yue Yu \\
 School of Optoelectronic Engineering\\
  Xidian University\\
  Xi’an 710071, Shaanxi, China \\
  	\And
  Joanna Borowiec\\
 Department of Electronic and Electrical Engineering\\
  University College London\\
  London, WC1E 7JE \\
  	\And
  Francisco V Ramirez-Cuevas \\
  Faculty of Engineering and Sciences\\
  Adolfo Ibáñez University\\
  Peñalolén, Santiago \\
  \And
   Danhong Yang \\
 Department of Mechanical Engineering\\
  University College London\\
  London, WC1E 7JE \\
  \And
   Souvik Ghosh \\
 Department of Electronic and Electrical Engineering\\
  University College London\\
  London, WC1E 7JE \\
   \And
  Cameron Tropea \\
 Institute for Fluid Mechanics and Aerodynamics\\
  Technische Universität Darmstadt\\
  64287 Darmstadt, Germany \\
  \And
     Ivan P. Parkin \\
 Chemistry Department\\
  University College London\\
  London, WC1H 0AJ \\ 
  \And
	\href{https://orcid.org/0000-0000-0000-0000}{\includegraphics[scale=0.06]{orcid.pdf}\hspace{1mm}Ioannis Papakonstantinou} \\
	Department of Electronic and Electrical Engineering\\
	University College London\\
	London, WC1E 7JE \\
	\texttt{i.papakonstantinou@ucl.ac.uk} \\
}

\renewcommand{\shorttitle}{\textit{arXiv} Template}

\hypersetup{
pdftitle={A template for the arxiv style},
pdfsubject={q-bio.NC, q-bio.QM},
pdfauthor={David S.~Hippocampus, Elias D.~Striatum},
pdfkeywords={Radiation Progapation, Inverse Diffusion Approximation, More},
}

\begin{document}
\maketitle

\begin{abstract}
	\textcolor{black}{Photon transport through a diffusing slab can be described by the radiative transfer equation (RTE). When the slab is highly scattering and weakly absorbing, the RTE simplifies to the diffusion equation. In this paper, an inverse diffusion approximation (IDA) method is numerically developed to determine the optical properties (reduced scattering $\mu'_s$ and absorption coefficient $\mu_a$) of a homogeneous slab using simple reflectance and transmission measurements with a spectrometer.}\vphantom{To guide the fabrication of radiative cooling samples, which should be designed to be highly scattering in the solar spectrum, we propose using the inverse diffusion approximation method (IDA) to extract the optical properties(reduced scattering $\mu'_s$ and absorption coefficients $\mu_a$) from a diffusing slab.}  The reflectance and transmission of a diffusing slab with an arbitrary thickness can then be predicted by solving the forward problem by using the calculated $\mu'_s$ and $\mu_a$. Our method is validated both numerically, by directly comparing with Monte-Carlo simulations, and experimentally, by comparing with measurements on ZnO/PDMS and TiO$_2$/PDMS composite polymer films with varying thicknesses. The IDA method is also applied to distinguish between different types of tissue. Overall, our method could be used to guide the design of radiative cooling reflectors, or lighting and optical display diffusers for applications in medical imaging and other fields.
  \vphantom{To validate this IDA method, we first compared the simulated reflectance and transmission between the diffusion approximation and the Monte Carlo method. This comparison was made for a scenario involving a Polydimethylsiloxane (PDMS) matrix embedded with ZnO particles, and the results showed excellent agreement. Further numerical validation has been conducted to compare the inverse-calculated $\mu'_s$ and $\mu_a$ with the rigorously calculated coefficients obtained using the Lorentz-Mie theory, \vphantom{The$\mu_a$ spans a wide range from $10^{-3}\mu m^{-1}$ to $10^{-18}\mu m^{-1}$,}and the inversely calculated results exhibit remarkable agreement with the exact values. Experimental validation has been conducted to predict the reflectance and transmission of TiO$_2$/PDMS matrix and ZnO/PDMS matrix. This was achieved by using the inversely calculated optical properties derived from a 1mm thick TiO$_2$/PDMS matrix and 1mm thick ZnO/PDMS matrix. The predicted reflectance and transmission agree well with the measured values. \textcolor{blue}{The absorption and scattering coefficients of the tissue are closely related to its structural properties. An experiment was conducted to explore the use of IDA in distinguishing between printing papers of different colors}. The inversely calculated $\mu_a$ corresponds well with the color appearance of the printing papers. }
\end{abstract}

% keywords can be removed

\section{Introduction}
Visible and near-infrared light transport in diffusing (or scattering) media plays significant role in numerous applications including life sciences, materials science, optical systems and others\cite{mandal2018hierarchically} \cite{steelman2019light}\cite{klose2002optical}\cite{klose2010forward}\cite{gorpas2010three}. For example, light transport has been extensively  studied in medical imaging to detect small inhomogeneities or morphological changes in human organs, which can indicate potential diseases\cite{chance1995optical}\cite{hebden1994enhanced}. In other cases, light diffusing films have been widely employed in radiative cooling applications, in lighting systems and in optical displays \cite{ouyang2016light}\cite{zhao2019radiative}\cite{yin2020terrestrial}\cite{fan2022photonics}\cite{kim2005pmma}\cite{guo2015light}. Photon propagation through scattering media can  be accurately described using electromagnetic wave theory. \textcolor{black}{Rigorous computational methods to calculate electromagnetic wave scattering include the finite difference time domain (FDTD) method\cite{taflove1987finite}, the extended boundary condition method (also known as T-Matrix method)\cite{waterman1965matrix} \cite{mishchenko2016first}\cite{wriedt1998review}, the discrete dipole approximation(DDA) \cite{draine1994discrete} \cite{yurkin2007discrete}}, and others. \textcolor{black}{However, these methods restrict the size of the computational domain to a maximum of several hundred wavelengths} \cite{yurkin2011discrete}. Beyond this limit, the computation becomes extremely time consuming, making it unsuitable for simulating the much larger objects often encountered in practice.

Light diffusion can alternatively be calculated by solving the radiative transfer equation (RTE), also known as the Boltzmann equation \cite{lenoble1985radiative}\cite{wang2007biomedical}, which accurately accounts for absorption and scattering caused by inhomogeneities within a host medium. However, due to its complexity, an analytical solution for the RTE can only be obtained in a few simple cases. As a result, the RTE is typically solved numerically using methods like Monte-Carlo statistical ray tracing \cite{stegmann2016comparison}\cite{li2019simulation}. A particular challenge arises in Monte-Carlo simulations when the concentration of scattering centers is high. This significantly reduces the optical mean free path, resulting in more frequent scattering events. To accurately account for the fate of these photons, the threshold for scattering events before rays are stopped needs to be increased. However, this may lead to unreasonable simulation times. Setting the threshold too low, on the other hand, may speed up simulations at the expense of considerable errors. As we show in the appendix (Table \ref{Simulation of reflectance and transmission of a porous PDMS matrix by varying the volume fraction of air pores}), when simulating air pores embedded in a polymer matrix, about 3\% of rays become trapped and are prematurely stopped for air pore volume fractions as low as 20\%. To put this in perspective, in radiative cooling, where high reflectance is desirable, it is not uncommon for the volume fraction of scattering centers to exceed 70\% and in some cases even 90\%. In this case, a large number of photons are trapped internally and erroneously accounted for as absorbed, resulting in lower reflectance values than the ones measured experimentally. 

In diffuse media with high albedo, the radiance becomes nearly isotropic after multiple scattering events. In such cases, it is common for the radiance to be approximated by an average intensity plus a diffuse flux vector\cite{ishimaru1978wave}. Consequently, the RTE can be reduced to a simpler diffusion equation\cite{kienle1997improved}, called the diffusion approximation (DA), which is easier to solve analytically or numerically. \textcolor{black}{Using the DA method, the thickness of the slab must be at least several transport lengths, over which the photon propagation direction becomes randomized \cite{ishimaru1978wave}. }Implementing the diffusion approximation in numerical simulations is generally more straightforward compared to the more complex doubling-adding method, which, however, tends to be applicable to a wider range of optical systems \cite{hulst1963new}\cite{prahl1995adding}. 

Schmitt et al. developed a Green function formulation for the diffusion equation that satisfy the boundary conditions for a slab geometry, enabling the determination of the steady state reflectance of human skin \cite{schmitt1990multilayer}. \textcolor{black}{In that work, the forward problem was solved, calculating the reflectance $R$ and transmittance $T$ of a diffusing slab as functions of the reduced scattering coefficient $\mu'_s$, which represents the net effect of scattering that alters the direction of photon transport and the absorption coefficient $\mu_a$, which measures the rate at which light is absorbed as it travels through the medium.} Our objective here is to solve the inverse problem, determining $\mu'_s$ and $\mu_a$ by measuring the reflectance $R$ and transmission $T$ of a diffuse slab using simple experimental techniques, such as a UV-Vis spectrometer. Since the formulations for $R$ and $T$ are non-linear with respect to $\mu'_s$ and $\mu_a$, the iterative Newton method is employed to solve for the functionals $\mu'_s(R,T)$ and $\mu_a(R, T)$\cite{wang2012gauss}. 
\section{Method}
\subsection{Formulation of the reflectance and transmission using diffusion approximation}
\label{sec:headings}
\begin{figure}[H]
\centering
\includegraphics[scale=0.72]{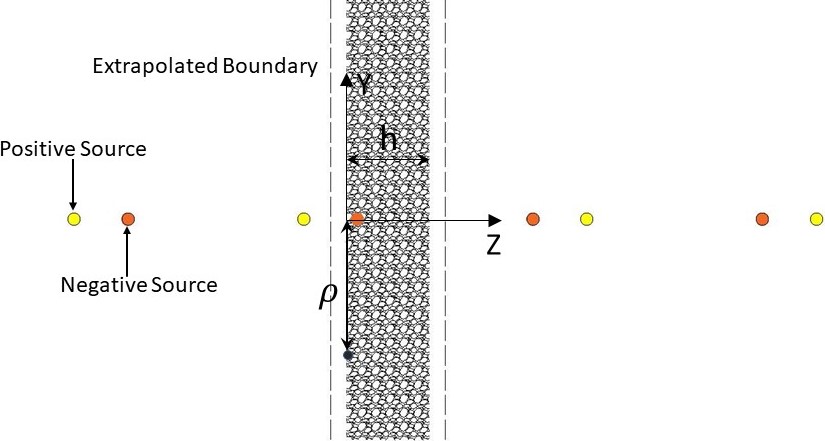}
\caption{Illustration of the extrapolated boundary and the positions of positive and negative dipoles to satisfy the boundary condition for a scattering medium. The dashed line represents the extrapolated boundary where the net diffusive flux is zero. The yellow circles represent the positive point sources and the orange circles represent the negative point sources.}
\label{The_extrapolated_boundary_and_the_positions_of_dipoles}
\end{figure}
The diffusion approximation is a simplification of the radiative transfer equation, and a detailed derivation can be found in the Supplementary Equations (S1)–(S19)\cite{ishimaru1978wave}\cite{groenhuis1983scattering}. Using the diffusion approximation method, the reflectance $R$ at a location $\rho(x,y)$ on the XY plane could be represented using Fick's law at the upper physical boundary of the medium, given as \cite{donner2006rapid}:
\begin{equation}
    R(\rho) = -D\nabla\phi(\rho, z = 0) = \sum_{i=-\infty}^{\infty}(\frac{\mu'_s}{4\pi (\mu'_s + \mu_a)}(\frac{z_{p,i}(1 + r_{p,i}\mu_{eff})e^{-\mu_{eff}r_{p,i}}}{r_{p,i}^3} - \frac{z_{n,i}(1 + r_{n,i}\mu_{eff})e^{-\mu_{eff}r_{n,i}}}{r_{n,i}^3}))
    \label{Formulation of reflectance at a point using the diffusion approximation method}
\end{equation}
where $\mu_{_{eff}}$ is the effective attenuation coefficient, and $D$ is the diffusion coefficient, given by:
\begin{equation}
D = \frac{1}{3(1 - g)\mu_s + \mu_a} = \frac{1}{3(\mu'_s + \mu_a)}
\end{equation}
\begin{equation}
\mu_{eff} = \sqrt{\frac{\mu_a}{D}}
\end{equation}
In the above equations $g$ is the anisotropy factor, and $\mu'_s$ is defined as $(1 - g)\mu_s$. The positions of the dipole sources are calculated by:
\begin{equation}
z_{p,i} = 2i(h + 2z_b) + z_0
\end{equation}
\begin{equation}
z_{n,i} = 2i(h + 2z_b) - z_0 - 2z_b
\end{equation}
\begin{equation}
r_{p,i} = \sqrt{\rho^2 + z_{p,i}^2}
\end{equation}
\begin{equation}
r_{n,i} = \sqrt{\rho^2 + z_{n,i}^2}
\end{equation}
\begin{equation}
\rho = \sqrt{x^2 + y^2}
\end{equation}
where\vphantom{$\rho$ is the distance between the light incident point of light source and the measured position in Y direction,} $h$ is the thickness of the slab, $z_0 = 1/(\mu_a + (1 - g)\mu_s)$, $z_b = 2AD$ and $i = 0,\pm 1,\pm 2,\pm 3\cdots$. $z_{p,i}$ and $z_{n,i}$ represent the locations of positive and negative sources correspondingly, as shown in Figure \ref{The_extrapolated_boundary_and_the_positions_of_dipoles}. In the diffusion approximation method, the first positive source must be located within the medium; otherwise, the calculated transmission may result in negative values. This occurs because both the positive and negative sources would then be positioned outside the slab boundaries.
Transmission can be calculated in a similar way by using Fick's law at the lower physical boundary of the medium, given as:
\begin{equation}
    T(\rho) = -D\nabla\phi(\rho, z = h) = \sum_{i=-\infty}^{\infty}(\frac{\alpha'}{4\pi}(\frac{z_{p,i}(1 + r_{p,i}\mu_{eff})e^{-\mu_{eff}r_{p,i}}}{r_{p,i}^3} - \frac{z_{n,i}(1 + r_{n,i}\mu_{eff})e^{-\mu_{eff}r_{n,i}}}{r_{n,i}^3}))
    \label{Formulation of transmission at a point using the diffusion approximation method}
\end{equation}
Finally, total reflectance and transmission are given by integration of $R(\rho)$ and $T(\rho)$ over the entire physical upper and lower boundaries respectively.
\begin{equation}
    R = \iint_S R(\rho)\mathrm{d}S 
    \label{Formulation of total reflectance}
\end{equation}
\begin{equation}
    T = \iint_S  T(\rho)\mathrm{d}S
    \label{Formulation of total transmission}
\end{equation}
\subsection{Inverse procedure for calculation of the reduced scattering coefficient and absorption coefficient}
\label{Inverse Diffusion Approximation}
\begin{figure}[H]
\centering
\includegraphics[scale=0.66]{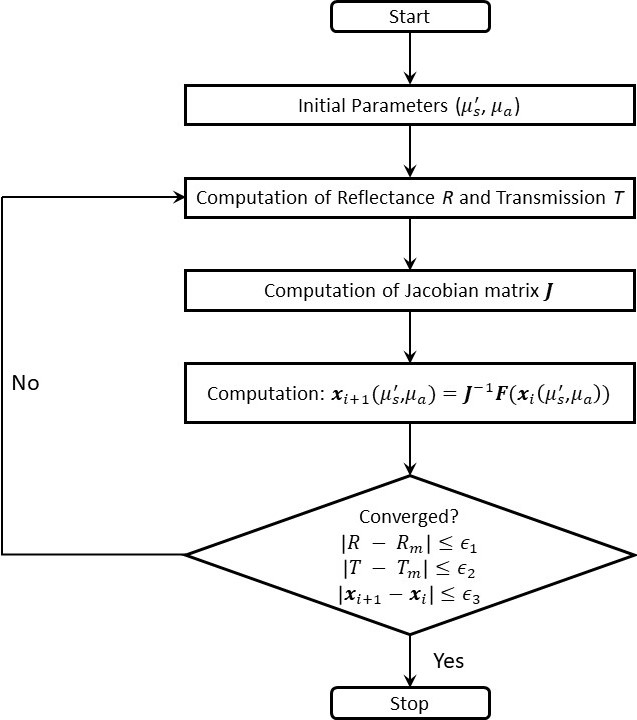}
\caption{Flow Chart for the Reconstruction of the Optical Properties}
\label{Flow Chart for the Reconstruction of the Optical Properties}
\end{figure}

In the formulation for the reflectance $R$ and transmission $T$ using the diffusion approximation method, both $R$ and $T$ are functions of $\mu'_s$ and $\mu_a$. The aim is to compute $\mu'_s$ and $\mu_a$ from measured total reflectance $R_m$ and transmission $T_m$ values for a diffusing slab geometry. The Gauss-Newton method has been used for the computation, according to the following equation:
\begin{equation}
    \mathbf{x_{n+1}}(\mu'_s,\mu_a) = \boldsymbol{J}^{-1}\boldsymbol{F}(\mathbf{x_{n}}(\mu'_s,\mu_a))
    \label{Newton Method}
\end{equation}
where the vector $\mathbf{x} = (\mu'_s \quad \mu_a)'$, $\boldsymbol{F}$ is the objective function, formulated in Equation (\ref{objective Function}), and $\boldsymbol{J}$ is the Jobobian Matrix, given by:
\begin{equation}
\boldsymbol{J} = 
    \begin{bmatrix}
        \frac{\partial R}{\partial \mu'_s} & \frac{\partial R}{\partial \mu_a}\\
        \frac{\partial T}{\partial \mu'_s} & \frac{\partial T}{\partial \mu_a}
\end{bmatrix}
    \label{Jacobian Matrix}
\end{equation}
\textcolor{black}{For the Jacobian matrix $\boldsymbol{J}$, an explicit expression for its matrix elements, representing the partial derivatives of $R$ and $T$, cannot be derived.} However, since the integrals in equation (\ref{Formulation of total reflectance}) and \ref{Formulation of total transmission} do not depend on $\mu'_s$ and $\mu_a$, the partial differential operator could be moved within the integral operator according to Leibniz's integral rule:
\begin{equation}
\boldsymbol{J} = 
    \begin{bmatrix}
        \iint_S \frac{\partial R(\rho)}{\partial \mu'_s}\mathrm{d}S  &\iint_S  \frac{\partial R(\rho)}{\partial \mu_a} \mathrm{d}S \\
        \iint_S \frac{\partial T(\rho)}{\partial \mu'_s}\mathrm{d}S & \iint_S \frac{\partial T(\rho)}{\partial \mu_a}\mathrm{d}S
\end{bmatrix}
    \label{Reformulation of Jacobian Matrix}
\end{equation}
In this case, the objective function $\boldsymbol{F}$ is given by:
\begin{equation}
\boldsymbol{F} = 
    \begin{bmatrix}
        R(\mathbf{x}) - R_m\\
        T(\mathbf{x}) - T_m
\end{bmatrix}
    \label{objective Function}
\end{equation}
in which $R_m$ and $T_m$ represent the measured values for total reflectance and transmittance, respectively. The procedure for extraction of $\mu'_s$ and $\mu_a$is illustrated in Figure \ref{Flow Chart for the Reconstruction of the Optical Properties}. At the beginning, an initial value for the reduced scattering coefficient and absorption coefficient is assigned. \textcolor{black}{In the current code, the initial values for $\mu'_s$ and $\mu_a$ are typically assigned starting from a small value, specifically $10^{-4} \cdot \mathrm{\mu m^{-1}}$ and $10^{-7} \cdot \mathrm{\mu m^{-1}}$ respectively, to ensure that the first multiplication with the Jacobian matrix yields a positive value.} Then the reflectance and transmission are computed using Equations (\ref{Formulation of total reflectance}) and (\ref{Formulation of total transmission}), as well as the Jocobian matrix. The calculated reflectance and transmission is then compared with the experimentally measured value. If the deviation between the calculated values and measured values are within a certain range, they are accepted; \textcolor{black}{otherwise, the newly computed value $\mathbf{x}_{i + 1}(\mu'_s,\mu_a)'$ will be used to calculate $R$ and $T$. This process is repeated until $\mathbf{x}(\mu'_s,\mu_a)'$, $R$ and $T$ converge.} 
\section{Numerical and Experimental Validation}
\label{Numerical and Experimental Validation}
The proposed method for utilizing the diffusion approximation to extract the reduced scattering coefficient $\mu'_s$ and $\mu_a$ has been validated both numerically and experimentally.
\subsection{Numerical Validation}
\begin{figure}[H]
\centering
\includegraphics[scale=0.75]{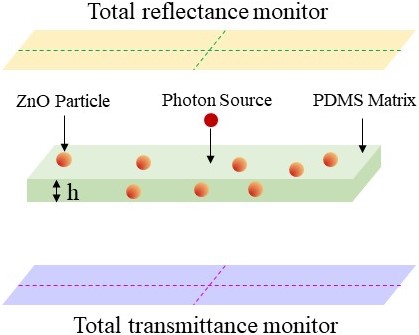}
\caption{Illustration for the scenario that ZnO particles are uniformly distributed within PDMS matrix. The size of the ZnO particles is 5 $\mu$m and thickness $h$ of the PDMS matrix is 1526 $\mu$m.}
\label{ZnO Particles within PDMS Matrix}
\end{figure}

Forward numerical validation was conducted to compare simulated reflectance and transmission using the Monte Carlo method with results calculated using the diffusion approximation method. This comparison was made for scenarios involving certain concentrations of ZnO particles uniformly embedded within a PDMS matrix, as illustrated in Figure \ref{ZnO Particles within PDMS Matrix}. Comparisons were made for two volume concentrations: 5\% and 10\% of ZnO particles within the PDMS matrix, and for an easy-to-fabricate thickness of 2 mm. \textcolor{black}{When good agreement is achieved for these two concentrations, the DA method should also work for higher volume concentrations, as their transport mean free path is even shorter.} At higher volume concentrations of ZnO particles, internal scattering increases significantly, causing a portion of the photons to become trapped within the matrix during simulation. For instance, at wavelength 550nm, the absorption coefficient of ZnO is zero, and the volume fraction of ZnO particles is set to 20\%, 3\% of the photons still remain unabsorbed and do not exit the matrix, when the threshold for the number of scattering events per single ray is set to 10,000. \textcolor{black}{These photons are classified as neither reflected or transmitted but rather erroneously as absorbed.}

\begin{figure}[H]
\centering
\includegraphics[scale=0.39]{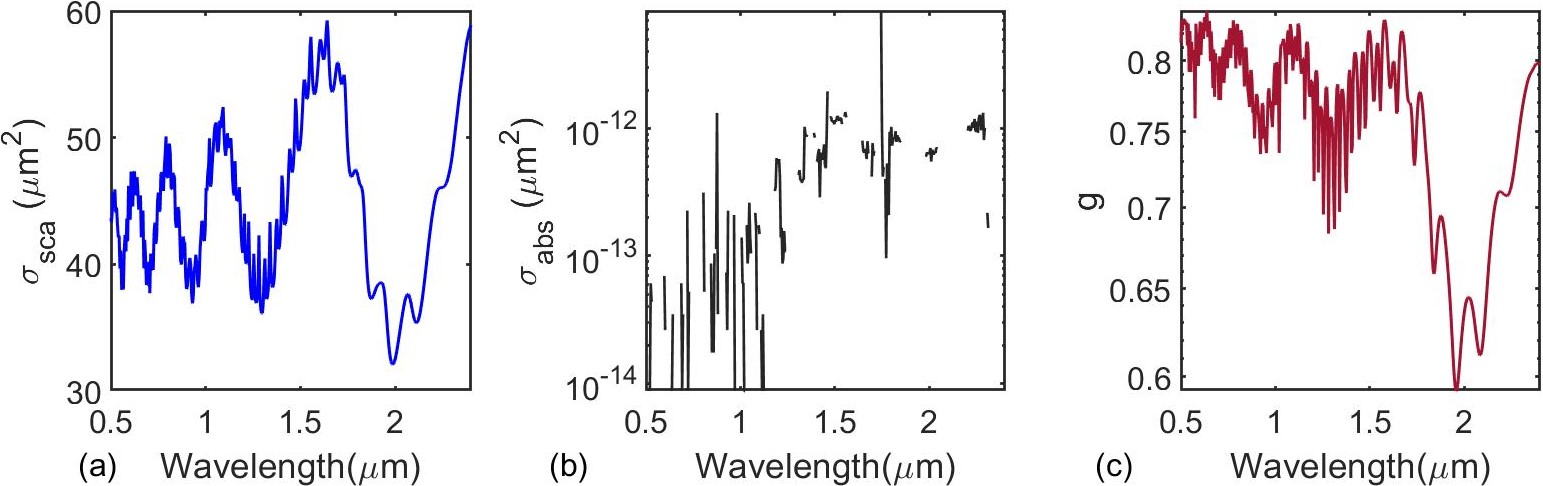}
\caption{Calculated scattering cross section $\sigma_{sca}$, absorption cross section $\sigma_{abs}$ and asymmetry parameter $g$ of a spherical ZnO particle using Lorenz-Mie theory. The diameter of the ZnO particle is 5 $\mu$m.}
\label{Calculated Optical Properties of a spherical ZnO particle}
\end{figure}

Prior to running the Monte-Carlo simulations, the scattering cross-section $\sigma_{sca}$, absorption cross-section $\sigma_{abs}$, and asymmetry parameter $g$  of the ZnO particles within the PDMS matrix were calculated using the Lorenz-Mie theory \cite{mie1908beitrage, bohren2008absorption}, as shown in Figure \ref{Calculated Optical Properties of a spherical ZnO particle} (a), (b) and (c). These parameters were then fed to our in-house developed Monte-Carlo simulator \cite{Ramirez2021}, where 100,000 normally incident photons were launched. The intensity of each individual photon transmitted (reflected) by the ZnO/PDMS matrix was then recorded. The total transmittance (reflectance) was then obtained by summing the intensities of all individual  photons. 

\begin{figure}[H]
\centering
\includegraphics[scale=0.33]{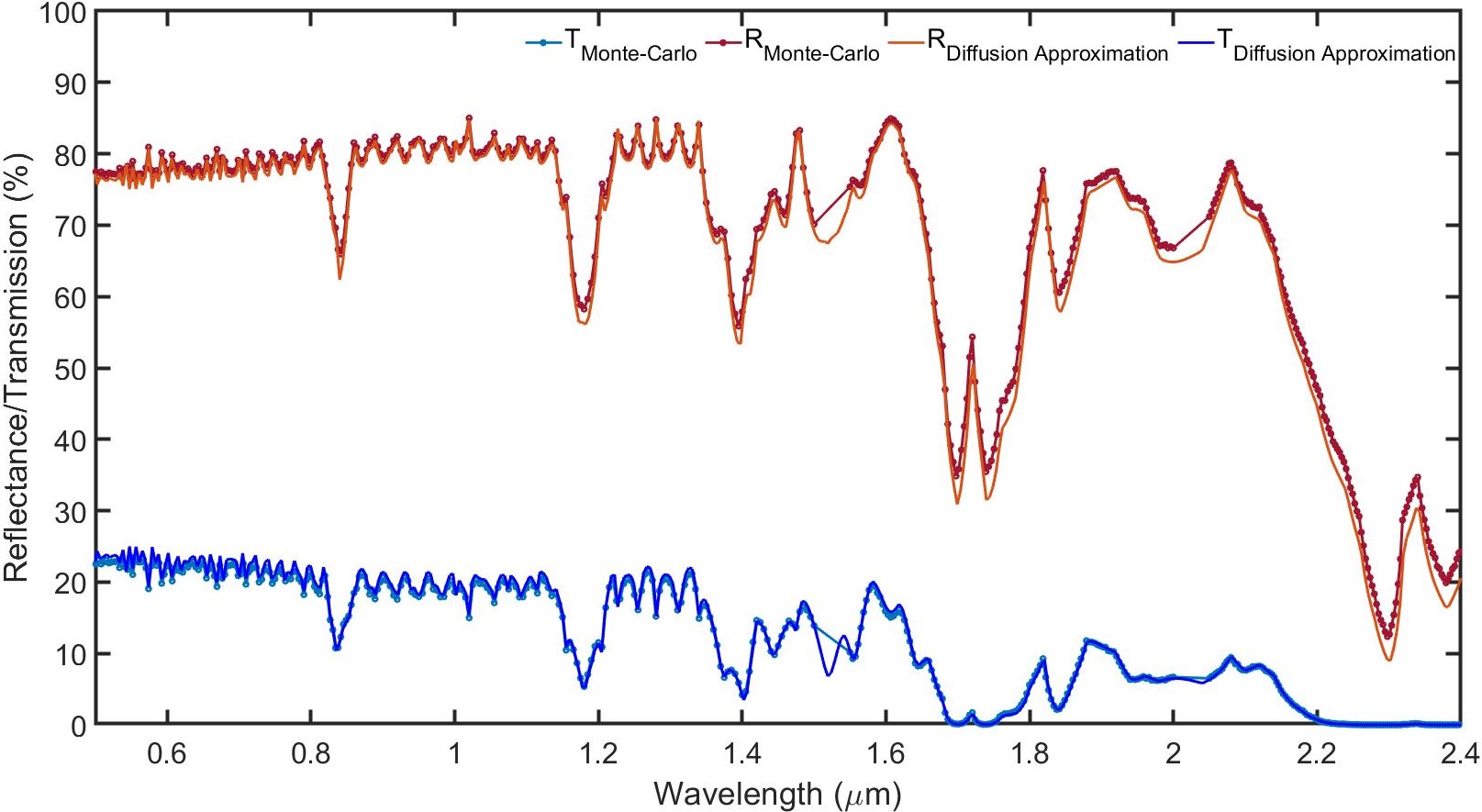}
\caption{Comparison of the simulated reflectance and transmission using the Monte-Carlo method with those calculated using the forward diffusion approximation method, for a ZnO volume concentration of 5\%. The orange and bright blue lines represent the calculated reflectance and transmission using the diffusion approximation method; the red and light blue lines with circular markers represent the simulated reflectance and transmission using the Monte-Carlo method.}
\label{Comparison between the the simulated reflectance and transmission using Monte-Carlo method and diffusion approximation method respectively}
\end{figure}

\begin{figure}[H]
\centering
\includegraphics[scale=0.33]{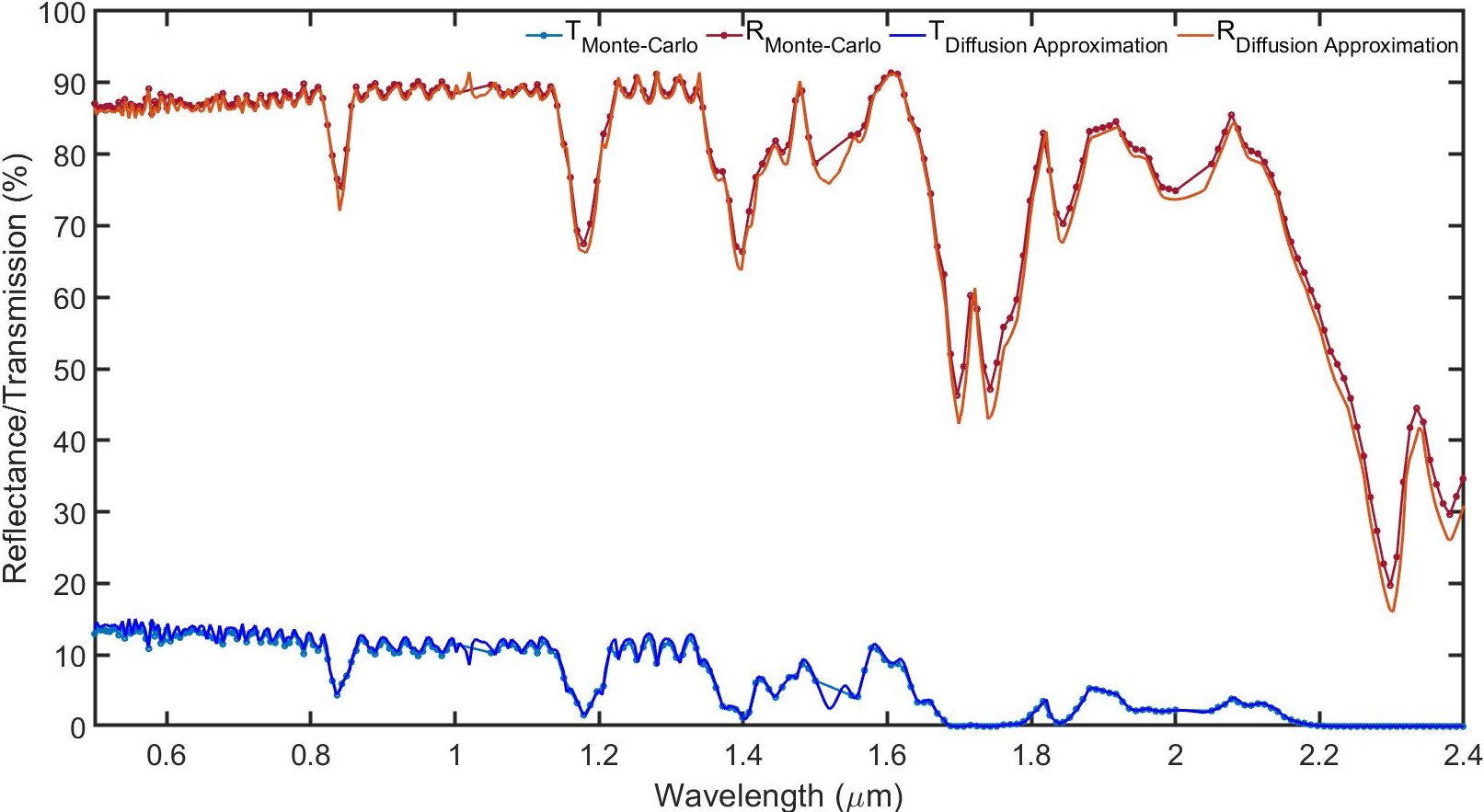}
\caption{Comparison of the simulated reflectance and transmission using the Monte-Carlo method with those calculated using the forward diffusion approximation method, for a ZnO volume concentration of 10\%. The orange and bright blue lines represent the calculated reflectance and transmission using the diffusion approximation method; the red and light blue lines with circular markers represent the simulated reflectance and transmission using the Monte-Carlo method.}
\label{Comparison 10 percent volume fraction}
\end{figure}

\begin{figure}[H]
\centering
\includegraphics[scale=0.45]{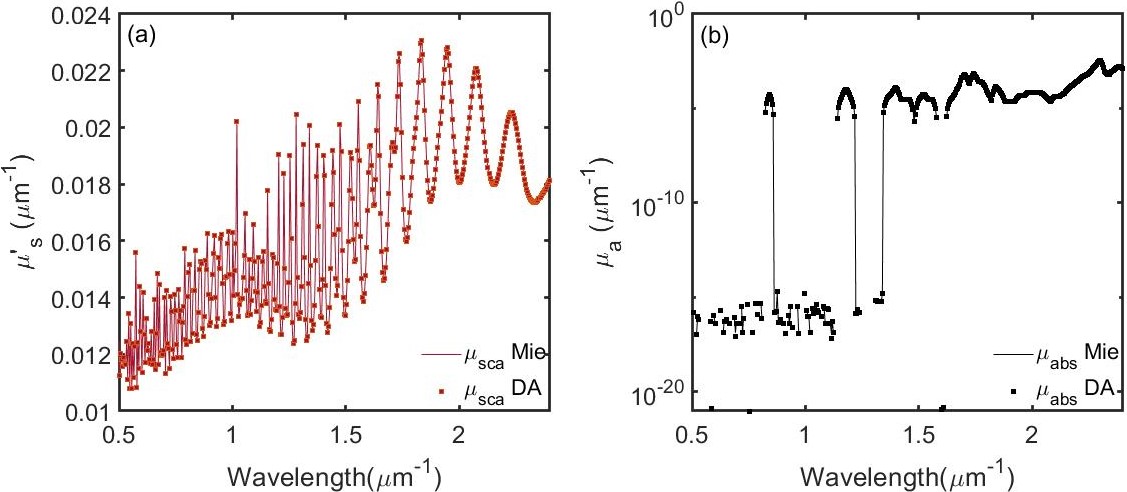}
\caption{Comparison of $\mu'_s$ and $\mu_a$ calculated using IDA method and Lorenz-Mie theory, when kowning the geometry and the values of reflectance $R$ and transmission $T$ of a ZnO/PDMS matrix}
\label{Validation Compare Inverse Results with Theoretical Value using Mie Theory}
\end{figure}

In the diffusion approximation method on the other hand, the values of $\mu'_s$ and $\mu_a$ need to be calculated. These are related to $\sigma_{sca}$, $\sigma_{abs}$ and $g$ via the following equations \cite{li2019simulation}:
\begin{equation}
    \mu'_s = (1 - g) \frac{V_f \sigma_{sca}}{V_p}
    \label{Formulation for reduced scattering coefficient}
\end{equation}
\begin{equation}
    \mu_a = \frac{V_f \sigma_{abs}}{V_p} + 4 \frac{\kappa}{\lambda}
    \label{Formulation for absorption coefficient}
\end{equation}
where, $V_f$ is the volume fraction of ZnO particles, $V_p$ is the volume of ZnO particle, $\kappa$ is the imaginary part of the refractive index of PDMS matrix and $\lambda$ is the wavelength. $R(\rho)$ and $T(\rho)$ are calculated using Equations (\ref{Formulation of reflectance at a point using the diffusion approximation method}) and (\ref{Formulation of transmission at a point using the diffusion approximation method}). The total reflectance and transmission have been calculated by integration of $R(\rho)$ and $T(\rho)$ over the entire upper and lower surface of the PDMS/ZnO matrix. The comparison results are shown in the Figure \ref{Comparison between the the simulated reflectance and transmission using Monte-Carlo method and diffusion approximation method respectively} and Figure \ref{Comparison 10 percent volume fraction} for the wavelength from 500 nm to 2.4 $\mu$m, the overall agreement between the diffusion approximation method and the Monte-Carlo simulations is excellent. 
A backward validation was performed in the following way: Using the reflectance $R$ and transmission $T$ presented in Figure \ref{Comparison between the the simulated reflectance and transmission using Monte-Carlo method and diffusion approximation method respectively}, $\mu'_s$ and $\mu_a$ where calculated using the IDA method described in section \ref{Inverse Diffusion Approximation} and compared against the values calculated by the Lorenz-Mie theory. The IDA calculated $\mu'_s$ and $\mu_a$ show excellent agreement with the Lorenz-Mie derived, as shown in Figure \ref{Validation Compare Inverse Results with Theoretical Value using Mie Theory}, which is remarkable, given that $\mu_a$ spans 15 orders of magnitudes, ranging from $10^{-3}\mu m^{-1}$ to $10^{-18}\mu m^{-1}$.
\subsection{Experimental Validation}
Experimental validation was carried out by comparing measured and calculated values for reflectance and transmission in various polymer composites of different thicknesses. We used PDMS as the host matrix, pigmented with  TiO$_2$ and ZnO particles. PDMS was chosen as it is a common polymer in many optical applications \cite{zhou2019polydimethylsiloxane}\cite{papakonstantinou2021hidden}. TiO$_2$ and ZnO are also commonly used pigments for light scattering due to their high refractive index\cite{chou2012preparation}. The diameter of the ZnO (Nanografi Nano Technology) particles varied between 30 - 50 nm, while the average size of TiO$_2$ (NanographeneX) was $\sim$200 nm. In both cases, the volume concentration of the pigments in the polymer matrix was set to 3\%, which guaranteed homogeneous mixing. The fabrication process is described in greater detail in Appendix \ref{Fabrication of TiO2/PDMS and ZnO/PDMS matrices},  and the ZnO/PDMS matrices are illustrated in Figure \ref{PDMS_ZnO_Matrix}.

\begin{figure}[H]
\centering
\includegraphics[scale=0.435]{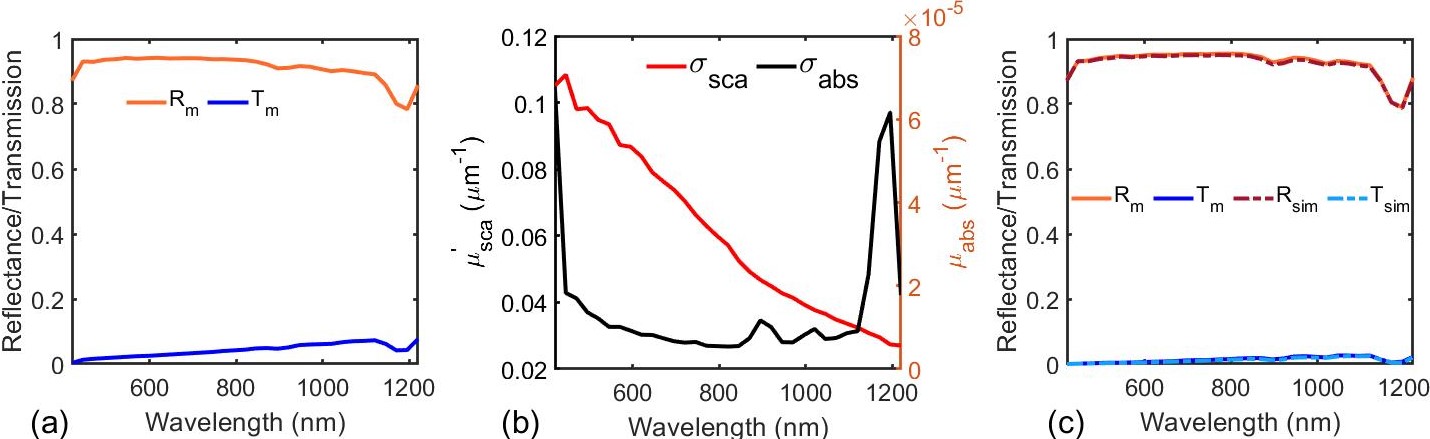}
\caption{Experimental validation of the IDA method by predicting the reflectance and transmission of a 2571$\mu$m thick PDMS/ZnO matrix using the inverse calculated $\mu'_s$ and $\mu_a$ values from a 912$\mu$m thick PDMS/ZnO matrix. Figure (a) presents the measured reflectance and transmission of the 912$\mu$m thick PDMS/ZnO matrix obtained with a UV-Vis spectrometer. Figure (b) displays the inverse-calculated $\mu'_s$ and $\mu_a$ for the 912$\mu$m thick PDMS/ZnO matrix. Figure (c) compares the simulated and measured reflectance and transmission for the 2571$\mu$m thick PDMS/ZnO matrix.}
\label{Experiment validation using the 200nm TiO2 particle}
\end{figure}

\begin{figure}[H]
\centering
\includegraphics[scale=0.435]{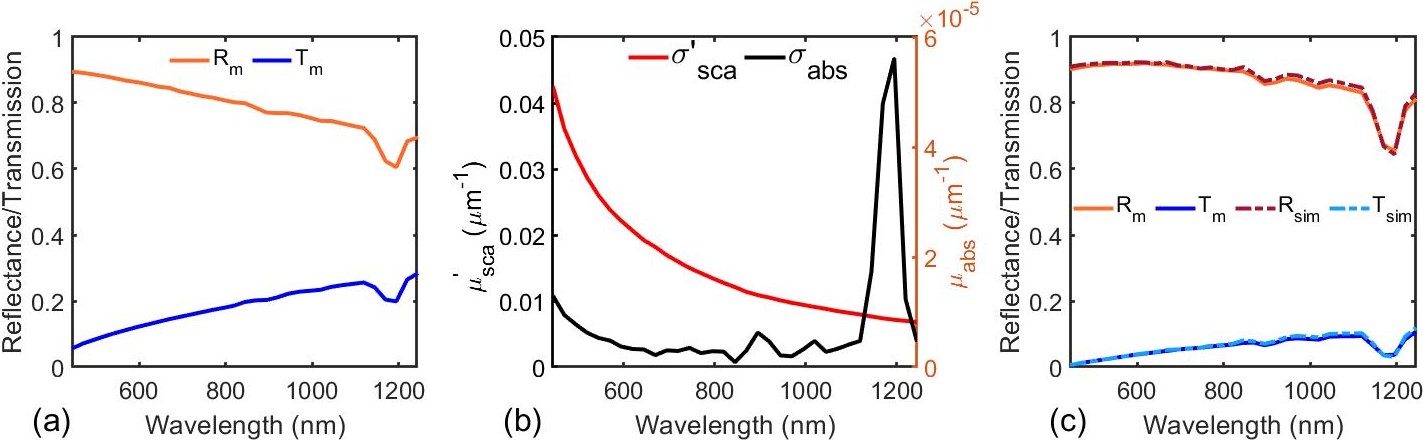}
\caption{Experimental validation of the IDA method by predicting the reflectance and transmission of a 3036$\mu$m thick PDMS/TiO$_2$ matrix using the inverse calculated $\mu'_s$ and $\mu_a$ values from a 943$\mu$m thick PDMS/TiO$_2$ matrix. Figure (a) presents the measured reflectance and transmission of the 912$\mu$m thick PDMS/TiO$_2$ matrix obtained with a UV-Vis spectrometer. Figure (b) displays the inverse calculated $\mu'_s$ and $\mu_a$ for the 943$\mu$m thick PDMS/TiO$_2$ matrix. Figure (c) compares the simulated and measured reflectance and transmission for the 3036$\mu$m thick PDMS/TiO$_2$ matrix.}
\label{Experiment validation}
\end{figure}

Initially, the  reflectance and transmission of $\sim$1 mm thick TiO$_2$/PDMS and ZnO/PDMS samples were measured in the range between 300 - 1300 nm using a UV-Vis-NIR spectrometer (Shimadzu UV-Vis 3600i). Results are shown in Figures \ref{Experiment validation using the 200nm TiO2 particle}(a) and \ref{Experiment validation}(a). The $\mu'_s$ and $\mu_a$ values for the two samples were then calculated using the IDA method, as shown in Figures \ref{Experiment validation using the 200nm TiO2 particle}(b) and \ref{Experiment validation}(b). Using the computed $\mu'_s$ and $\mu_a$ values, the forward DA method was used to calculate the transmission/reflectance of $\sim$3 mm thick samples. As shown in Figures \ref{Experiment validation using the 200nm TiO2 particle}(c) and \ref{Experiment validation}(c), excellent agreement was obtained between experiments and simulations, confirming that the IDA method can be used to predict the optical properties of arbitrary thick samples, provided they all share the same scattering properties.    
%\begin{figure}[H]
%\centering
%\includegraphics[scale=0.5]{Figures/Integrating Sphere Light Path.jpg}
%\caption{Integrating Sphere Light Path of Shimadzu UV-3600i Plus}
%\label{Integrating Sphere Light Path}
%\end{figure}

\begin{figure}[H]
\centering
\includegraphics[scale=0.45]{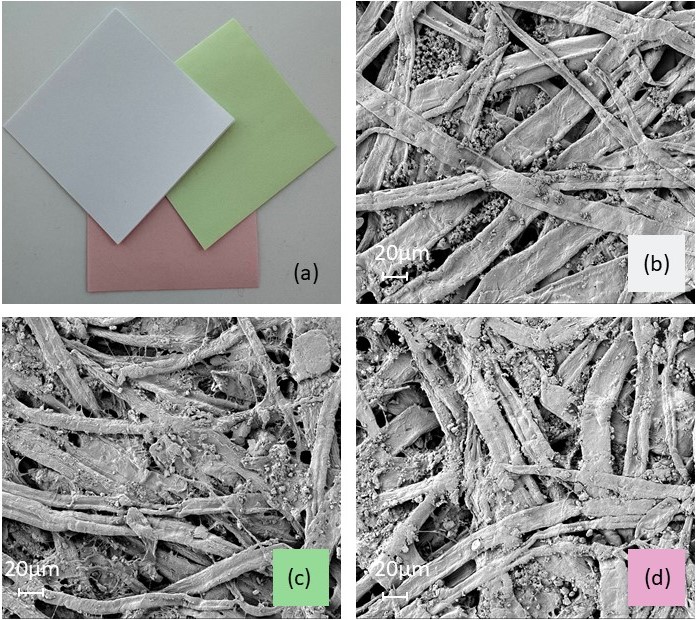}
\caption{(a) Printing papers with white, pink and green-cyan colors respectively (b) SEM images of printing papers in white, green, and pink, revealing their internal fiber structure.}
\label{Printing Paper}
\end{figure}

\begin{figure}[H]
\centering
\includegraphics[scale=0.38]{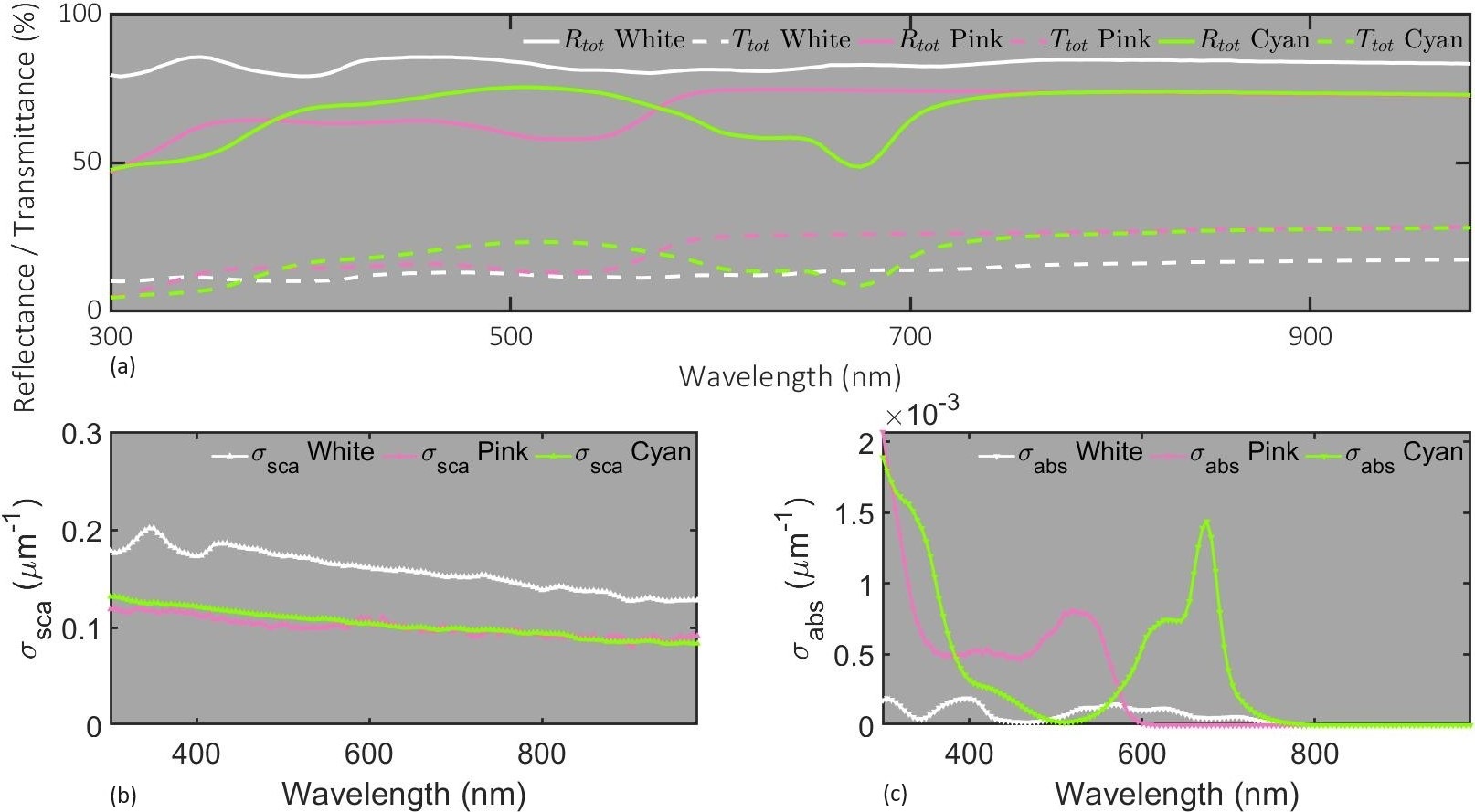}
\caption{Measured reflectance and transmission of printing papers using a Shimadzu UV-Vis 3600i plus spectrometer. In (a), the solid white, pink, and cyan lines represent the measured reflectance $R$ for white, pink, and cyan printing paper, respectively. The dashed lines represent their measured transmission $T$. (b) and (c) shows the reconstructed $\mu'_s$ and $\mu_a$ values using the measured $R$ and $T$.}
\label{Reconstruction_Optical_Properties_White_Pink_Blue_Print_Papers}
\end{figure}

A second experiment was conducted to investigate the effectiveness of IDA in distinguishing between different tissues. For this study, white, pink, and green-cyan colored printing papers were used, as shown in Figure \ref{Printing Paper} (a). The internal structure of these materials was examined using scanning electron microscopy, as shown in Figure \ref{Printing Paper} (b), (c), and (d). The original SEM images can be found in the supplementary Figures \ref{SEM_image_for_green_print_paper}, \ref{SEM_image_for_red_print_paper}, and \ref{SEM_image_for_white_print_paper}. The thickness of each paper was measured $\sim$ 93 $\mu$m. The reflectance $R$ and transmission $T$ values of the papers were measured by UV-Vis-NIR and presented in Figure \ref{Reconstruction_Optical_Properties_White_Pink_Blue_Print_Papers} (a). The refractive index of these materials was unknown and was measured using spectrometric ellipsometry\cite{bakker2004determination}. Results from the reconstruction of $\mu'_s$ and $\mu_a$ are presented in Figure \ref{Reconstruction_Optical_Properties_White_Pink_Blue_Print_Papers}(b) and \ref{Reconstruction_Optical_Properties_White_Pink_Blue_Print_Papers}(c). \textcolor{black}{The pink and cyan printing papers exhibit a similar $\mu'_s$ when compared to the white printing paper, as both the pink and cyan papers are sourced from the same company and batch of product. The structural and composition differences among these papers are also evident in their individual SEM images \ref{SEM_image_for_green_print_paper}, \ref{SEM_image_for_red_print_paper}, and \ref{SEM_image_for_white_print_paper}, where the fibers of the white printing paper appear wider and smoother.}
The $\mu_a$ of white printing paper is lower compared to the $\mu_a$ of pink and green-cyan paper. Pink color, is made of mixing white with red, the latter extending between 625nm to 740nm. From the extracted $\mu_a$ of pink paper, we see that $\mu_a$ drops to very low  values starting from $\sim$ 600nm, as expected. The Cyan color consists mostly of green (495nm to 570nm) and blue (450nm to 495nm) components. The calculated $\mu_a$, nicely captures those too, being lower in this region compared to the rest of the visible wavelengths. 
\section{Discussion and Conclusion}
An inverse diffusion approximation (IAD) method has been developed to derive the wavelength dependent $\mu'_s$ and $\mu_a$ of a scattering medium from its total reflectance and transmission. \textcolor{black}{To achieve this, the Newton method was utilized, and the Jacobian matrix was derived.} This method underwent numerical and experimental validations. In forward numerical validation, the computed reflectance and transmission using the diffusion approximation method were compared with those simulated using the Monte-Carlo method, yielding very good agreement for a ZnO/PDMS matrix when the volume concentrations of ZnO particle are 5\% and 10\% respectively. \textcolor{black}{Further numerical validation was conducted when given the values of reflectance $R$ and transmission $T$, from which $\mu'_s$ and $\mu_a$ were calculated using the IAD method. These results were then compared with the theoretical values derived from Lorenz-Mie theory, showing excellent agreement.} To experimentally validate the IDA method, we used the $\mu'_s$ and $\mu_a$ values of ZnO/PDMS and TiO$_2$/PDMS matrices, calculated via IDA from reflectance ($R$) and transmission ($T$) data measured using a UV-Vis spectrometer, to predict the reflectance and transmission of ZnO/PDMS and TiO$_2$/PDMS samples with varying thicknesses over the wavelength range of 400 nm to 1300 nm. The diffusion approximation is applicable when the slab thickness is several times greater than the transport mean free path, which holds true within this wavelength range. At 1300 nm, the largest transport mean free path, approximately 100 micrometers, occurs for the PDMS/TiO$_2$ composite, while the smallest transport mean free path, around 9 micrometers, occurs for the PDMS/ZnO matrix at 400 nm. Therefore, the thicknesses of the PDMS matrices range from 9 to 300 times the transport mean free path, ensuring good agreement between the measured and predicted reflectance and transmission values. Since the absorption and scattering coefficients of tissue are closely tied to its structural characteristics. In an experiment aimed at evaluating the use of IDA for distinguishing between printing papers of different colors, the inversely calculated $\mu_a$ demonstrated a clear correlation with the observed color of the papers. \textcolor{black}{The IDA method can also be applied to evaluate the light scattering and absorption properties of materials like porous media, composite materials, and coatings. This information of $\mu'_s$ and $\mu_a$ is essential for optimizing the design of optical materials and controlling their transparency, diffuse reflectivity, or overall optical performance. For instance, it could be applied in the development of radiative cooling materials, which often incorporate porous structures or nanoparticles to enhance light scattering and improve total solar reflectance.}

However, while the IAD method shows promise, several limitations must be considered. The method is most effective under the assumption that scattering dominates over absorption. In cases where absorption is significant, the diffusion approximation may break down, leading to inaccuracies in the derived optical coefficients. Furthermore, the method may struggle with highly heterogeneous materials, where variations in particle size or distribution could lead to deviations from expected behavior as described by the diffusion approximation. Additionally, the thickness of the material must be at least several times greater than the transport mean free path for the diffusion approximation to be valid. Furthermore, the method may struggle with highly heterogeneous materials, where variations in particle size or distribution could lead to deviations from expected behavior as described by the diffusion approximation.
\section{Acknowledgement}
We would like to thank the UKRI, ERC proof-of-concept grant PolyCool EP/X024482/1 for funding.
\bibliographystyle{unsrt}
%\bibliography{references}  %%% Uncomment this line and comment out the ``thebibliography'' section below to use the external .bib file (using bibtex) .
\clearpage
\renewcommand\thefigure{\thesection.\arabic{figure}} 
\appendix
\setcounter{figure}{0} 
\counterwithin{figure}{section}
\section{Appendix}
\renewcommand{\theequation}{S.\arabic{equation}}
Photon transport within a medium can be described by the radiative transfer equation (RTE), which incorporates the law of conservation of energy:
\begin{equation}
	(\frac{1}{v\partial t} + \sigma_a + \sigma_s + \hat{\mathbf{s}} \cdot \mathbf{\nabla}) I(\mathbf{r},\hat{\mathbf{s}},t) = \sigma_t \iint_{4\pi} I(\mathbf{r},\hat{\mathbf{s}}^{'},t)f(\hat{\mathbf{s}},\hat{\mathbf{s}}^{'})\mathrm{d}\Omega + Q(\mathbf{r},\hat{\mathbf{s}}^{'},t) 
\end{equation}
\vphantom{In which $I(\mathbf{r},\hat{\mathbf{s}},t)$ is the intensity in $W \cdot m^{-2}  sr^{-1}$, $v$ is the speed of light within the medium; $\sigma_a$, $\sigma_s$ and $\sigma_t$ are the absorption coefficient, scattering coefficient and the extinction coefficient at $\mathbf{r}$ respectively. $\hat{\mathbf{s}}$ is the unit vector for the photon propagation direction. $Q(\mathbf{r},\hat{\mathbf{s}}^{'},t)$ is the source term. $f(\hat{\mathbf{s}},\hat{\mathbf{s}}^{'})$ is the scattering phase function, which is a probability density function for the photon transport in the direction $\hat{\mathbf{s}}^{'}$ being scattered into new direction $\hat{\mathbf{s}}$ into solid angle $\mathrm{d}\Omega$ after the scattering event at $\mathbf{r}$, $f(\hat{\mathbf{s}},\hat{\mathbf{s}}^{'})$ usually does not capture the interference effect.}
in which $I(\mathbf{r},\hat{\mathbf{s}},t)$ is the \textcolor{black}{specific} intensity in $\mathrm{W \cdot m^{-2}  sr^{-1}}$; $\sigma_a$, $\sigma_s$ and $\sigma_t$ are the absorption coefficient, scattering coefficient and the extinction coefficient at $\mathbf{r}$ respectively. $\hat{\mathbf{s}}$ is the unit vector for the photon propagation direction. $Q(\mathbf{r},\hat{\mathbf{s}}^{'},t)$ is the source term. $f(\hat{\mathbf{s}},\hat{\mathbf{s}}^{'})$ is the scattering phase function, which is a probability density function for the photon transport in the direction $\hat{\mathbf{s}}^{'}$ being scattered into new direction $\hat{\mathbf{s}}$ into solid angle $\mathrm{d}\Omega$ after the scattering event at $\mathbf{r}$. $f(\hat{\mathbf{s}},\hat{\mathbf{s}}^{'})$ usually does not capture the interference effect between the scatters.
$g$ is the anisotropy factor, representing the averaged probability of the forward scattering, defined as:
\begin{equation}
g = \frac{\iint_{4\pi}f(\hat{\mathbf{s}},\hat{\mathbf{s}}^{'}) (\hat{\mathbf{s}}\cdot \hat{\mathbf{s}}^{'})\mathrm{d}\Omega}{\iint_{4\pi}f(\hat{\mathbf{s}},\hat{\mathbf{s}}^{'})\mathrm{d}\Omega}           \;  \; \; \; \; \;  g \in [-1,1]
\end{equation}
For pure forward scattering, $g$ should be 1.
In which $v$ is the speed of light within the medium
\begin{equation}
	\iint_{4\pi}f(\hat{\mathbf{s}},\hat{\mathbf{s}}^{'})\mathrm{d}\Omega = \frac{\sigma_s}{\sigma_t}
\end{equation}
For the steady state, the time dependent term $\frac{1}{v\partial t}$  can be eliminated, and the RTE can be simplified as:
\begin{equation}
	(\sigma_a + \sigma_s + \hat{\mathbf{s}} \cdot \mathbf{\nabla}) I(\mathbf{r},\hat{\mathbf{s}}) = \sigma_t \iint_{4\pi} I(\mathbf{r},\hat{\mathbf{s}}')f(\hat{\mathbf{s}},\hat{\mathbf{s}}^{'})\mathrm{d}\Omega + Q(\mathbf{r},\hat{\mathbf{s}}^{'})
 \label{Steady State Radiative Transfer equation}
\end{equation}
The intensity can be represented using a spherical harmonics expansion.
\begin{equation}
I(\mathbf{r},\hat{\mathbf{s}}) = \sum_{l=0}^{n}\sum_{m = -n}^{m = n} I_l(\mathbf{r}) Y_{l}^m(\hat{\mathbf{s}})
\end{equation}
in which $Y_{l}^m(\hat{\mathbf{s}})$ denotes the spherical harmonics. In a diffusing medium with a high scattering albedo (\textcolor{black}{$\mu_a \ll \mu'_s$ and usually the scattering coefficient is at least one order larger than the absorption coefficient}\cite{wang2007biomedical}), it's assumed that light scatters nearly uniformly in all directions, resulting in an almost uniform angular distribution of the flux. The radiance $I(\mathbf{r},\hat{\mathbf{s}})$ can be expressed as a combination of isotropic term flux rate and a minor flux vector\cite{ishimaru1978wave}. Using the orthogonality properties of spherical harmonics, factoring out the isotropic scattering term and diffuse term, and neglecting the remaining terms\cite{ishimaru1978wave}, we can obtain:
\begin{equation}
I(\mathbf{r},\hat{\mathbf{s}}) = \frac{1}{4\pi}\phi(\mathbf{r}) +  \frac{3}{4\pi}\mathbf{j}(\mathbf{r})
\label{Formulation for diffuse specific intensity}
\end{equation}

By defining:
\begin{equation}
	\phi(\mathbf{r}) = 	\iint_{4\pi} I(\mathbf{r},\hat{\mathbf{s}}^{'})\mathrm{d}\Omega 
\end{equation}
\begin{equation}
	\mathbf{j}(\mathbf{r}) = 	\iint_{4\pi} I(\mathbf{r},\hat{\mathbf{s}}^{'}) \cdot \hat{\mathbf{s}} \; \mathrm{d}\Omega 
\end{equation}
$\phi(\mathbf{r})$ denotes the fluence rate, while $\mathbf{j}(\mathbf{r})$ is the diffuse flux vector, representing the net energy flux in the direction $\hat{\mathbf{s}}$. Integrating Equation (\ref{Steady State Radiative Transfer equation}) over the entire solid angle, we obtain\cite{ishimaru1978wave}:
\begin{equation}
	\mathbf{\nabla} \cdot \mathbf{j}(\mathbf{r}) = -\mu_a \phi(\mathbf{r}) + S(\mathbf{r})
 \label{Formulation for divergence of radiance flux vector}
\end{equation}
\begin{equation}
	S(\mathbf{r}) = 	\iint_{4\pi} Q(\mathbf{r})\mathrm{d}\Omega 
\end{equation}
% After expanding the phase function with the Legendre polynomials, and take the first two terms. The coefficient $a_l$ can be obtained by using the orthogonality properties of the Legendre Polynomials.
% \begin{equation}
% f(\hat{\mathbf{s}},\hat{\mathbf{s}}^{'}) = \sum_{l=1}^{n} a_l P_{l}(x)
% \end{equation}
% $P_{l}(x)$ is the Legendre function and $a_l=  (2l+1)/2 \int_{-1}^{1}f(\hat{\mathbf{s}},\hat{\mathbf{s}}^{'}) p_l (\cos{\theta})\mathrm{d} \cos{\theta})$. Then the $a_0=1$ and $a_1=3g$. 
Substituting Equation (\ref{Formulation for diffuse specific intensity}) into Equation (\ref{Steady State Radiative Transfer equation}), then multiplying both sides by $\hat{\mathbf{s}}$ and again integrating it over the entire solid angle yields
\begin{equation}
	(\mu_a + (1-g)\mu_s) \mathbf{j}(\mathbf{r}) = -\frac{1}{3}\phi(\mathbf{r})
 \label{Relationship between fluence rate and diffuse flux vector}
\end{equation}
Using Equations (\ref{Formulation for divergence of radiance flux vector}) and (\ref{Relationship between fluence rate and diffuse flux vector}), the steady state diffusion equation can be obtained:
\begin{equation}
	\mu_a\phi(\mathbf{r}) - D\nabla^2\phi(\mathbf{r}) = S(\mathbf{r})
 \label{Steady state diffusion equation}
\end{equation}
To obtain the particular solution for the diffusion equation, boundary conditions are required. The commonly used boundary conditions are zero boundary condition(ZBC), extrapolated boundary condition and partial current boundary condition(PCBC)\cite{kienle1997improved}. The extrapolated boundary condition has been used, which assumes that the average diffuse intensity at an extrapolated boundary outside the medium at a distance $z_b$ is zero. Compared with ZBC, this boundary condition has been reported to give a more accurate result when a mismatch of the refractive index at the boundary exists\cite{kienle1997improved}\cite{contini1997photon}. The boundary condition is formulated as following:
\begin{equation}
\iint_{\hat{\mathbf{s}} \cdot \hat{\mathbf{n}} > 0 }I(\mathbf{r},\hat{\mathbf{s}}) (\hat{\mathbf{s}},\mathbf{r}) \mathrm{d}\Omega= \iint_{\hat{\mathbf{s}} \cdot \hat{\mathbf{n}} < 0} R(\hat{\mathbf{n}})I(\mathbf{r},\hat{\mathbf{s}})(-\hat{\mathbf{s}},\mathbf{r}) \mathrm{d}\Omega
\end{equation}
in which $\hat{\mathbf{n}}$ is the normal vector at the interface, and $R(\hat{\mathbf{n}})$ is the Fresnel reflection coefficient for unpolarized incident wave\cite{contini1997photon}. Substituting Equation (\ref{Formulation for diffuse specific intensity}) in the above boundary condition, we obtain:

\begin{equation}
    \phi(\mathbf{r}) + \frac{A}{2\pi}\mathbf{j}(\mathbf{r}) \cdot \hat{\mathbf{n}}  = 0
    \label{Reformulation of Boundary Condition}
\end{equation}
and A is given by\cite{contini1997photon}:
\begin{equation}
    A = \frac{1 + 3\int_{0}^{\pi / 2} R(\theta_i)\cos^2{\theta_i}\sin{\theta_i}\mathrm{d}\theta_i}{1 - 2\int_{0}^{\pi / 2} R(\theta_i)\cos{\theta_i}\sin{\theta_i}\mathrm{d}\theta_i}
\end{equation}
Using the Fick's law ( $\mathbf{j}(\mathbf{r}) = -D\nabla \phi(\mathbf{r})$) in Equation (\ref{Reformulation of Boundary Condition}), we obtain:
\begin{equation}
\phi(\mathbf{r}) - 2AD \hat{\mathbf{n}} \cdot \nabla \phi(\mathbf{r}) = 0
\label{Partial Boundary Condition}
\end{equation}

Invoking the Green's function, the solution of the diffusion equation can be obtained. For a short-pulsed point source in an infinite homogeneous medium, the solution has the form\cite{donner2006rapid}:
\begin{equation}
    \phi(\mathbf{r}) = \frac{1}{4\pi Dr}e^{-\mu_{eff}r}
    \label{Solution of the diffusion equation for a point source within an infinite medium}
\end{equation}
in which $\mu_{eff}= \sqrt{\frac{\mu_a}{D}}$ and $r= \lvert \mathbf{r}  - \mathbf{r'}\lvert$, denoting the distance between the position where the fluence rate is measured and the position of the source. For the extropolated boundary condtion that the fluence rate becomes zero at a distance beyond the actual surface $z_e = 2AD$, for a homogeneous semi-infinite medium, the solution to satisfy the boundary conditions:
\begin{equation}
    \phi(\mathbf{r}) = \frac{1}{4\pi D}(\frac{e^{-\mu_{eff}r_{p}}}{r_{p}} - \frac{e^{-\mu_{eff}r_{n}}}{r_{n}})
    \label{Solution of the diffusion equation for a point source within a homogeneous semi-infinite medium}
\end{equation}
For a homogeneous diffusing slab, the boundary conditions can be satisfied with infinite pairs of positive and negative point sources\cite{schmitt1990multilayer}:
\begin{equation}
    \phi(\mathbf{r}) = \sum_{i=-\infty}^{\infty}(\frac{1}{4\pi D}(\frac{e^{-\mu_{eff}r_{pi}}}{r_{pi}} - \frac{e^{-\mu_{eff}r_{ni}}}{r_{ni}}))
    \label{Solution of the diffusion equation for a point source within a homogeneous slab}
\end{equation}

\section{Monte-Carlo Simulation}
Radiative transfer simulations were conducted using the open source software MC-Photon, which is designed for Monte Carlo simulations of unpolarized light. \cite{Ramirez2021}. The code has been carefully validated previously\cite{ramirez2022universal}. The Monte Carlo algorithm simulates the trajectories of numerous individual photons as they interact with particles and interfaces until they are either absorbed or leave the simulation domain. The initial conditions for each photon are defined by the position and direction of the light source. At each simulation step, the optical path $\lambda_{photon}$ and the fate of a photon are determined by selecting the shortest path among the particle scattering $\lambda_{sca}$, absorption $\lambda_{abs}$, the absorption of the host $\lambda_{host}$, or diffraction $\lambda_{Fresnel}$, where:
\begin{equation}
	\lambda_{sca} = \frac{V_p}{f_v \bigl \langle C_{sca}\bigr \rangle}  \ln(\xi) 
\end{equation}
\begin{equation}
	\lambda_{sca} = \frac{V_p}{f_v \bigl \langle C_{abs}\bigr \rangle}  \ln(\xi)
\end{equation}
\begin{equation}
	\lambda_{host} = \frac{\lambda}{4\pi\kappa_{host}}  \ln(\xi)
\end{equation}
and $\lambda_{Fresnel}$ is given by the shortest distance between the photon and an interface. In the equations above, $\lambda$ is the wavelength, $\xi$ is a random number between 0 and 1, $\kappa_{host}$ is the imaginary part of the refractive index of the host, $v_P$ is the volume of a particle and $f_v$ is the volume fraction of particles within the host.
If diffraction occurs ($\lambda_{photon} = \lambda_{Fresnel}$), the photon is either reflected or transmitted, determined by random selection. The probabilities of these events are proportional to the respective energy fluxes as defined by Fresnel's equations. If the photon is absorbed by a particle ($\lambda_{photon} = \lambda_{abs}$) or the host material ($\lambda_{photon} = \lambda_{host}$), the event is terminated, and the simulation proceeds with a new photon starting from the initial conditions. For a scattered photon (($\lambda_{photon} = \lambda_{sca}$)) the new direction, $\theta$, is determined by:
\begin{equation}
	\cos\theta = \begin{cases}
\frac{1}{2g} [1 + g^2 - (\frac{1-g^2}{1 - g + 2g\xi})^2] \quad if \quad g \neq 0 \\
2\xi - 1 \quad if \quad g = 0.
\end{cases}
\end{equation}
where $g = \bigl \langle \mu_{sca} \bigr \rangle$, and $\mu_{sca}$  is the asymmetry parameter.
In all simulations, a slab with a large surface area was used to represent a 2D problem. The smallest surface area was selected as a criterion to ensure that no photons escape through the edges. Two large detectors placed above and below the slab measure the total reflectance and transmittance, respectively. Each simulation considered 1,000,000 photons per wavelength.

The reflectance and transmission of a porous PDMS matrix, used in the development of a prototype for radiative cooling coatings, were simulated by varying the volume fraction of air pores from 1\% to 50\%. The dimensions of the PDMS matrix are 50 mm in length, 50 mm in width, and 2 mm in thickness, with an air pore size of 100 nm. For each simulation, 1,000,000 photons were launched to ensure statistical convergence of the results. The threshold for scattering events for each ray is set to 10,000 interactions. The table below presents the percentage of trapped photons as the volume fraction of air pores within the PDMS matrix increases, at a wavelength of 550 nm.
\begin{table}[H]
\begin{center}
\caption{Simulation of reflectance and transmission of a porous PDMS matrix by varying the volume fraction of air pores from 1\% to 50\%.}
\label{Simulation of reflectance and transmission of a porous PDMS matrix by varying the volume fraction of air pores}
\begin{tabular}{||c c c c||} 
 \hline
 Volume Fraction of Air Pore & Simulated Reflectance &Simulated Transmission & Trapped Photon \\ [0.5ex] 
 \hline\hline
 1 & 0.783 & 0.217 & 0 \\ 
 \hline
 5 & 0.944 & 0.056 & 0 \\
 \hline
 10 & 0.967 & 0.0263 & 0.0067 \\
 \hline
 20 & 0.969 & 0.0022 & 0.0288 \\
 \hline
 50 & 0.969 & 0 & 0.0310 \\ [1ex] 
 \hline
\end{tabular}
\end{center}
\end{table}
\section{Fabrication of \texorpdfstring{TiO$_{2}$ / PDMS and}  Z ZnO / PDMS matrices}\label{Fabrication of TiO2/PDMS and ZnO/PDMS matrices}
\begin{figure}[H]
\centering
\includegraphics[scale=0.12]{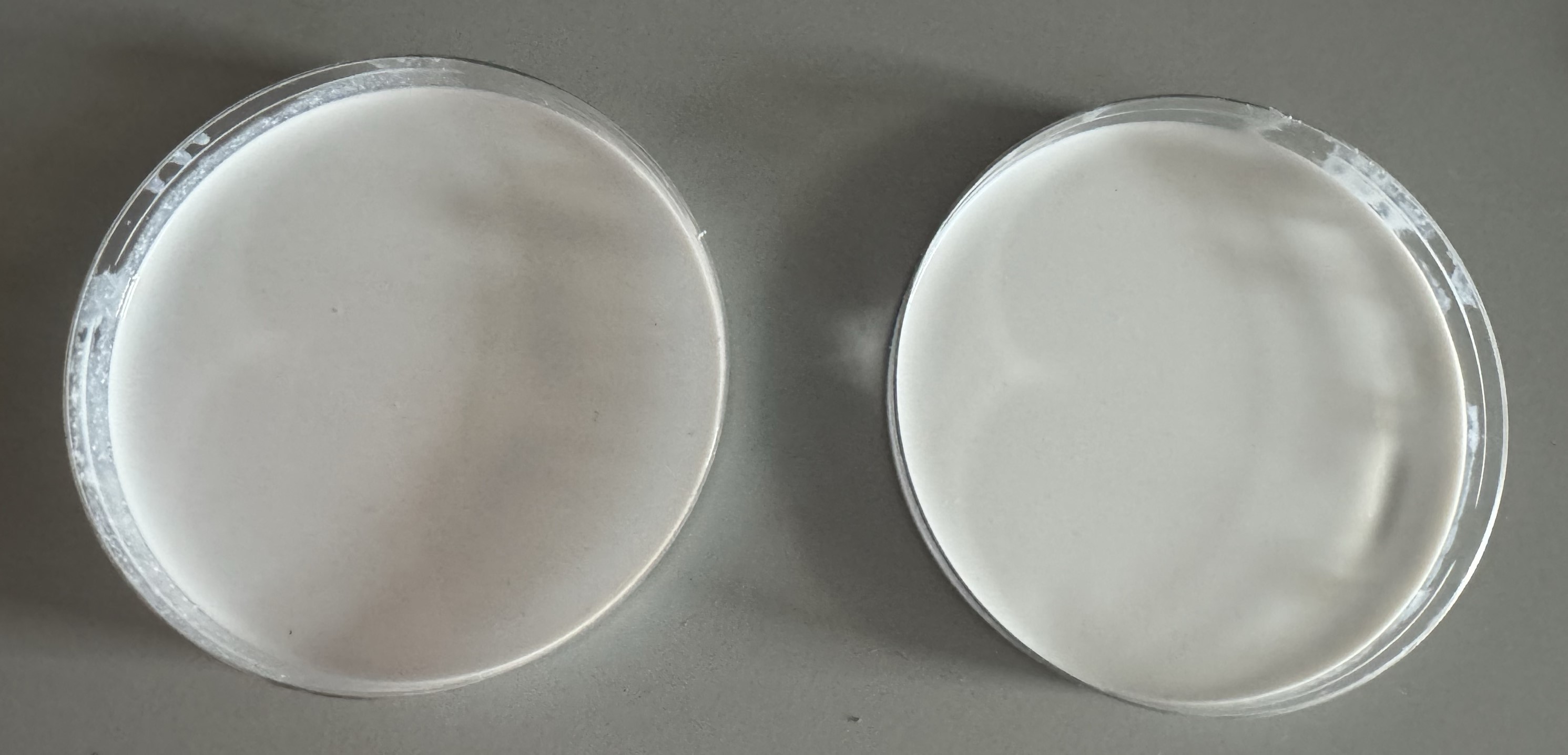}
\caption{PDMS ZnO matrix. The left matrix is 1 mm thick, and the right matrix is 3 mm thick.}
\label{PDMS_ZnO_Matrix}
\end{figure}

 TiO$_2$/PDMS and ZnO/PDMS matrices were prepared to validate the IDA model. PDMS (polydimethylsiloxane, Sylgard 184, Dow Corning) was ordered from Dow Chemical. It also contains a curing agent. When the curing agent is mixed with PDMS at a 1:10 weight ratio, the PDMS will solidify. Before pouring the TiO$_2$/PDMS and ZnO/PDMS liquid mixtures into molds for solidification, the mixtures were placed in a Femto Plasma cleaner to remove air bubbles. After degassing, the mixtures were poured into molds to solidify. The molds were then placed on a hot plate set to 50 degrees Celsius to accelerate the solidification process. The solidified matrices are shown in Figure \ref{PDMS_ZnO_Matrix}.

\section{SEM Images for Printing Papers}
\begin{figure}[H]
\centering
\includegraphics[scale=0.25]{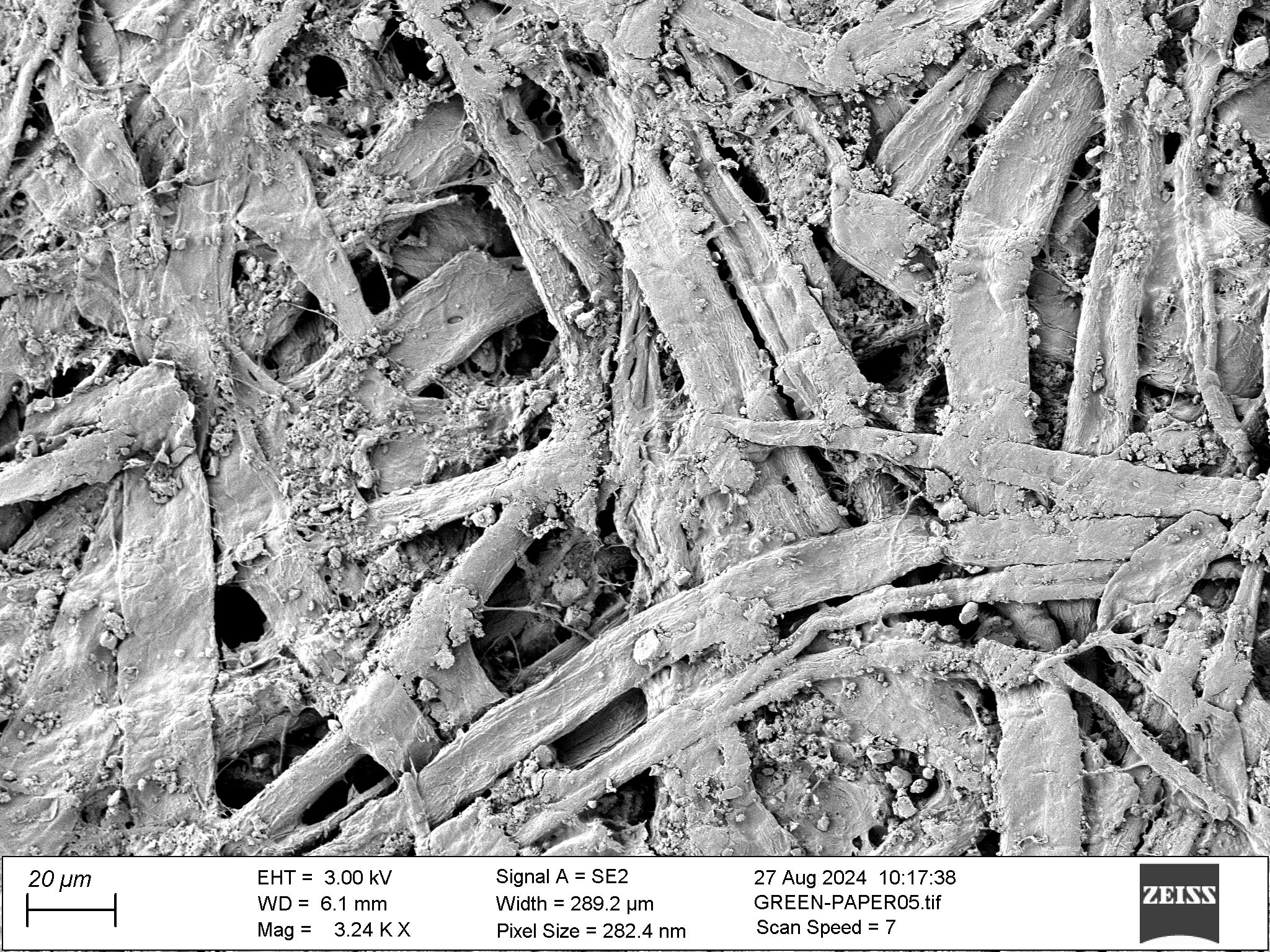}
\caption{SEM image for green print paper.}
\label{SEM_image_for_green_print_paper}
\end{figure}

\begin{figure}[H]
\centering
\includegraphics[scale=0.25]{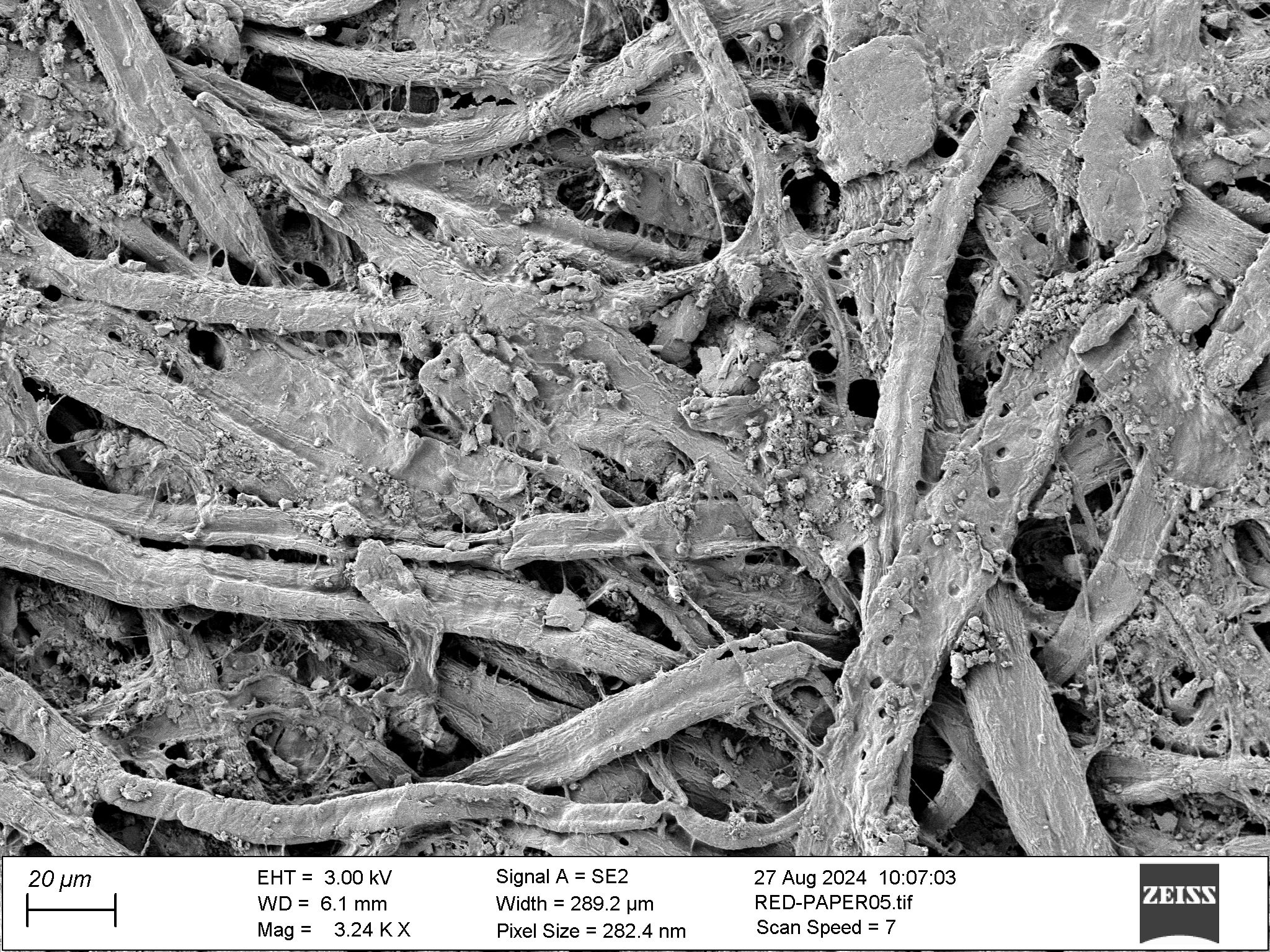}
\caption{SEM image for red print paper.}
\label{SEM_image_for_red_print_paper}
\end{figure}

\begin{figure}[H]
\centering
\includegraphics[scale=0.25]{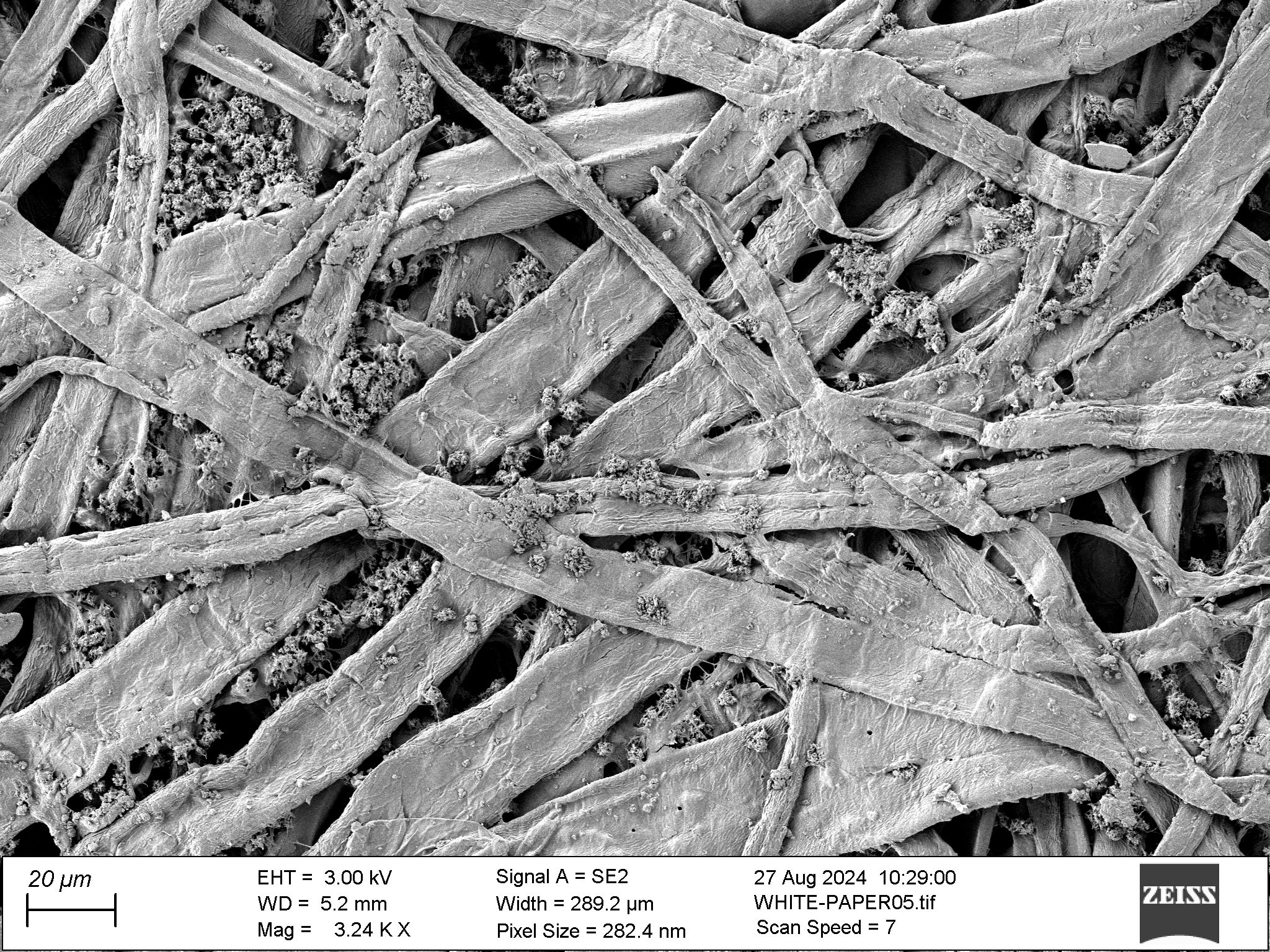}
\caption{SEM image for white print paper.}
\label{SEM_image_for_white_print_paper}
\end{figure}


\begin{thebibliography}{10}

\bibitem{mandal2018hierarchically}
Jyotirmoy Mandal, Yanke Fu, Adam~C Overvig, Mingxin Jia, Kerui Sun, Norman~N Shi, Hua Zhou, Xianghui Xiao, Nanfang Yu, and Yuan Yang.
\newblock Hierarchically porous polymer coatings for highly efficient passive daytime radiative cooling.
\newblock {\em Science}, 362(6412):315--319, 2018.

\bibitem{steelman2019light}
Zachary~A Steelman, Derek~S Ho, Kengyeh~K Chu, and Adam Wax.
\newblock Light-scattering methods for tissue diagnosis.
\newblock {\em Optica}, 6(4):479--489, 2019.

\bibitem{klose2002optical}
Alexander~D Klose, Uwe Netz, J{\"u}rgen Beuthan, and Andreas~H Hielscher.
\newblock Optical tomography using the time-independent equation of radiative transfer—part 1: forward model.
\newblock {\em Journal of Quantitative Spectroscopy and Radiative Transfer}, 72(5):691--713, 2002.

\bibitem{klose2010forward}
Alexander~D Klose.
\newblock The forward and inverse problem in tissue optics based on the radiative transfer equation: a brief review.
\newblock {\em Journal of Quantitative Spectroscopy and Radiative Transfer}, 111(11):1852--1853, 2010.

\bibitem{gorpas2010three}
Dimitris Gorpas, Dido Yova, and Konstantinos Politopoulos.
\newblock A three-dimensional finite elements approach for the coupled radiative transfer equation and diffusion approximation modeling in fluorescence imaging.
\newblock {\em Journal of Quantitative Spectroscopy and Radiative Transfer}, 111(4):553--568, 2010.

\bibitem{chance1995optical}
Britton Chance.
\newblock Optical tomography, photon migration, and spectroscopy of tissue and model media-theory, human studies and instrumentation.
\newblock In {\em Proc. SPIE}, volume 2389, 1995.

\bibitem{hebden1994enhanced}
Jeremy~C Hebden and David~T Delpy.
\newblock Enhanced time-resolved imaging with a diffusion model of photon transport.
\newblock {\em Optics Letters}, 19(5):311--313, 1994.

\bibitem{ouyang2016light}
Xing Ouyang, Peibin Li, Dazhu Chen, and Jiaoning Tang.
\newblock Light-diffusing materials for led illumination applications: Comparing the effectiveness of two typical light-diffusing agents.
\newblock {\em Journal of Applied Polymer Science}, 133(4), 2016.

\bibitem{zhao2019radiative}
Bin Zhao, Mingke Hu, Xianze Ao, Nuo Chen, and Gang Pei.
\newblock Radiative cooling: A review of fundamentals, materials, applications, and prospects.
\newblock {\em Applied Energy}, 236:489--513, 2019.

\bibitem{yin2020terrestrial}
Xiaobo Yin, Ronggui Yang, Gang Tan, and Shanhui Fan.
\newblock Terrestrial radiative cooling: Using the cold universe as a renewable and sustainable energy source.
\newblock {\em Science}, 370(6518):786--791, 2020.

\bibitem{fan2022photonics}
Shanhui Fan and Wei Li.
\newblock Photonics and thermodynamics concepts in radiative cooling.
\newblock {\em Nature Photonics}, 16(3):182--190, 2022.

\bibitem{kim2005pmma}
GeunHyung Kim.
\newblock A pmma composite as an optical diffuser in a liquid crystal display backlighting unit (blu).
\newblock {\em European Polymer Journal}, 41(8):1729--1737, 2005.

\bibitem{guo2015light}
Shuang Guo, Shuxue Zhou, Huijing Li, and Bo~You.
\newblock Light diffusing films fabricated by strawberry-like pmma/sio2 composite microspheres for led application.
\newblock {\em Journal of Colloid and Interface Science}, 448:123--129, 2015.

\bibitem{taflove1987finite}
A~Taflove and KR~Umashankar.
\newblock The finite-difference time-domain (fd-td) method for electromagnetic scattering and interaction problems.
\newblock {\em Journal of Electromagnetic Waves and Applications}, 1(3):243--267, 1987.

\bibitem{waterman1965matrix}
Peter~C Waterman.
\newblock Matrix formulation of electromagnetic scattering.
\newblock {\em Proceedings of the IEEE}, 53(8):805--812, 1965.

\bibitem{mishchenko2016first}
Michael~I Mishchenko, Janna~M Dlugach, Maxim~A Yurkin, Lei Bi, Brian Cairns, Li~Liu, R~Lee Panetta, Larry~D Travis, Ping Yang, and Nadezhda~T Zakharova.
\newblock First-principles modeling of electromagnetic scattering by discrete and discretely heterogeneous random media.
\newblock {\em Physics Reports}, 632:1--75, 2016.

\bibitem{wriedt1998review}
Thomas Wriedt.
\newblock A review of elastic light scattering theories.
\newblock {\em Particle \& Particle Systems Characterization: Measurement and Description of Particle Properties and Behavior in Powders and Other Disperse Systems}, 15(2):67--74, 1998.

\bibitem{draine1994discrete}
Bruce~T Draine and Piotr~J Flatau.
\newblock Discrete-dipole approximation for scattering calculations.
\newblock {\em JOSA A}, 11(4):1491--1499, 1994.

\bibitem{yurkin2007discrete}
Maxim~A Yurkin and Alfons~G Hoekstra.
\newblock The discrete dipole approximation: an overview and recent developments.
\newblock {\em Journal of Quantitative Spectroscopy and Radiative Transfer}, 106(1-3):558--589, 2007.

\bibitem{yurkin2011discrete}
Maxim~A Yurkin and Alfons~G Hoekstra.
\newblock The discrete-dipole-approximation code adda: Capabilities and known limitations.
\newblock {\em Journal of Quantitative Spectroscopy and Radiative Transfer}, 112(13):2234--2247, 2011.

\bibitem{lenoble1985radiative}
Jacqueline Lenoble.
\newblock {\em Radiative Transfer in Scattering and Absorbing Atmospheres: Standard Computational Procedures}, volume 300.
\newblock A. Deepak Hampton, Va., 1985.

\bibitem{wang2007biomedical}
Lihong~V Wang and Hsin-i Wu.
\newblock {\em \textnormal{Biomedical Optics: Principles and Imaging}}.
\newblock John Wiley \& Sons, New Jersey, 2007.

\bibitem{stegmann2016comparison}
Patrick~G Stegmann, Cameron Tropea, Emma J{\"a}rvinen, and Martin Schnaiter.
\newblock Comparison of measured and computed phase functions of individual tropospheric ice crystals.
\newblock {\em Journal of Quantitative Spectroscopy and Radiative Transfer}, 178:379--389, 2016.

\bibitem{li2019simulation}
Lingxi Li, Patrick~G Stegmann, Simon Rosenkranz, Walter Sch{\"a}fer, and Cameron Tropea.
\newblock Simulation of light scattering from a colloidal droplet using a polarized monte carlo method: application to the time-shift technique.
\newblock {\em Optics Express}, 27(25):36388--36404, 2019.

\bibitem{ishimaru1978wave}
Akira Ishimaru.
\newblock {\em \textnormal{Wave Propagation and Scattering in Random Media}}, volume~2.
\newblock Academic Press, New York, 1978.

\bibitem{kienle1997improved}
Alwin Kienle and Michael~S Patterson.
\newblock Improved solutions of the steady-state and the time-resolved diffusion equations for reflectance from a semi-infinite turbid medium.
\newblock {\em JOSA A}, 14(1):246--254, 1997.

\bibitem{hulst1963new}
Hendrik~Christoffel Hulst.
\newblock {\em \textnormal{A new look at multiple scattering}}.
\newblock NASA Institute for Space Studies, Goddard Space Flight Center, 1963.

\bibitem{prahl1995adding}
Scott~A Prahl.
\newblock The adding-doubling method.
\newblock In {\em Optical-Thermal Response of Laser-Irradiated Tissue}, pages 101--129. Springer, 1995.

\bibitem{schmitt1990multilayer}
JM~Schmitt, GX~Zhou, EC~Walker, and RT~Wall.
\newblock Multilayer model of photon diffusion in skin.
\newblock {\em JOSA A}, 7(11):2141--2153, 1990.

\bibitem{wang2012gauss}
Yong Wang.
\newblock Gauss--{Newton} method.
\newblock {\em Wiley Interdisciplinary Reviews: Computational Statistics}, 4(4):415--420, 2012.

\bibitem{groenhuis1983scattering}
RAJ Groenhuis, Hedzer~A Ferwerda, and JJ~Ten~Bosch.
\newblock Scattering and absorption of turbid materials determined from reflection measurements. 1: Theory.
\newblock {\em Applied Optics}, 22(16):2456--2462, 1983.

\bibitem{donner2006rapid}
Craig Donner and Henrik~Wann Jensen.
\newblock Rapid simulation of steady-state spatially resolved reflectance and transmittance profiles of multilayered turbid materials.
\newblock {\em JOSA A}, 23(6):1382--1390, 2006.

\bibitem{mie1908beitrage}
Gustav Mie.
\newblock Beitr{\"a}ge zur {Optik} tr{\"u}ber {Medien}, speziell kolloidaler {Metall{\"o}sungen}.
\newblock {\em Annalen der Physik}, 330(3):377--445, 1908.

\bibitem{bohren2008absorption}
Craig~F Bohren and Donald~R Huffman.
\newblock {\em \textnormal{Absorption and Scattering of Light by Small Particles}}.
\newblock John Wiley \& Sons, New York, 2008.

\bibitem{Ramirez2021}
F~Ramirez-Cuevas.
\newblock Mc-photon. a {Monte-Carlo} software for photon transport.
\newblock https://github.com/PanxoPanza/mc-photon.git, 2021.

\bibitem{zhou2019polydimethylsiloxane}
Lyu Zhou, Haomin Song, Jianwei Liang, Matthew Singer, Ming Zhou, Edgars Stegenburgs, Nan Zhang, Chen Xu, Tien Ng, Zongfu Yu, et~al.
\newblock A polydimethylsiloxane-coated metal structure for all-day radiative cooling.
\newblock {\em Nature Sustainability}, 2(8):718--724, 2019.

\bibitem{papakonstantinou2021hidden}
Ioannis Papakonstantinou, Mark Portnoi, and Michael~G Debije.
\newblock The hidden potential of luminescent solar concentrators.
\newblock {\em Advanced Energy Materials}, 11(3):2002883, 2021.

\bibitem{chou2012preparation}
Chuen-Shii Chou, Ming-Geng Guo, Kuan-Hung Liu, and Yi-Siang Chen.
\newblock Preparation of {TiO2} particles and their applications in the light scattering layer of a dye-sensitized solar cell.
\newblock {\em Applied Energy}, 92:224--233, 2012.

\bibitem{bakker2004determination}
JWP Bakker, G~Bryntse, and Hans Arwin.
\newblock Determination of refractive index of printed and unprinted paper using spectroscopic ellipsometry.
\newblock {\em Thin Solid Films}, 455:361--365, 2004.

\bibitem{contini1997photon}
Daniele Contini, Fabrizio Martelli, and Giovanni Zaccanti.
\newblock Photon migration through a turbid slab described by a model based on diffusion approximation. {I}. theory.
\newblock {\em Applied Optics}, 36(19):4587--4599, 1997.

\bibitem{ramirez2022universal}
Francisco~V Ramirez-Cuevas, Kargal~L Gurunatha, Ivan~P Parkin, and Ioannnis Papakonstantinou.
\newblock Universal theory of light scattering of randomly oriented particles: a fluctuational-electrodynamics approach for light transport modeling in disordered nanostructures.
\newblock {\em ACS Photonics}, 9(2):672--681, 2022.

\end{thebibliography}




% \end{thebibliography}

\end{document}